\documentclass[aps,prb,reprint,showpacs,groupedaddress]{revtex4-2}

\usepackage{bm}
\usepackage{wasysym}
\usepackage{amssymb}
\usepackage[cp1251]{inputenc}
\usepackage[dvips]{graphicx}
\usepackage[usenames]{color}
\begin{document}

\title{Plasmons and magnetoplasmons in partially bounded two-layer electron systems\\}

\author{A.A. Zabolotnykh}
\affiliation{Kotelnikov Institute of Radio-engineering and Electronics of the RAS, Mokhovaya 11-7, Moscow 125009, Russia\\
}

\author{V.A. Volkov}
\email{Volkov.V.A@gmail.com}
\affiliation{Kotelnikov Institute of Radio-engineering and Electronics of the RAS, Mokhovaya 11-7, Moscow 125009, Russia\\
}

\date{\today}

\begin{abstract}
We have analytically studied plasmons in an electron system comprised of two spatially separated layers --- an infinite two-dimensional electron system (2DES) and a 2D strip. Our analysis reveals the existence of plasmon modes that are localized near and propagate along the strip. These modes are characterized by the wave vector in the direction of the strip, as well as the number of charge density nodes, $N$, across the strip. In the long-wavelength limit, the fundamental mode $N=0$ is found to have gapless linear dispersion. When the external perpendicular magnetic field is applied, this mode remains gapless and exhibits peculiar magnetodispersion. We analyze the correlation between our findings and the previously established results on plasmons in gated and partially gated 2DESs.
\end{abstract}

\maketitle

\section{Introduction}
Plasma oscillations, or plasmons, have been studied extensively in 2DESs since the pioneering work~\cite{Stern1967} was first reported over fifty years ago. Neglecting electromagnetic retardation effects, plasmons in a 2DES embedded in a dielectric medium with constant permittivity $\varkappa$ are described by the gapless square-root dispersion law:
\begin{equation}
  \label{plasmon}	
  \omega_{p}(q)=\sqrt{\frac{2\pi n e^2q}{\varkappa m}}, \quad q=\sqrt{q_x^2+q_y^2},
\end{equation} 
where $n$ is the 2D electron concentration, $m$ is the electron effective mass, and $q$ is the 2D wave vector of the plasmon. The derivation of the dispersion relation (\ref{plasmon}) relies on the assumption of infinite electron relaxation time and long-wavelength limit, $q\ll k_F$, where $\hbar k_F$ is the Fermi momentum.

Considering an infinite metal gate placed near and parallel to the 2DES, in the long-wavelength limit $qh\ll 1$, with $h$ being the distance between the gate and 2DES, the plasmon frequency softens by a factor of $\sqrt{2hq}$ leading to the linear dispersion of so-called gated plasmons~\cite{Chaplik1972}:
\begin{equation}
\label{gated}
	\omega_g(q)=qV_p,\quad V_p= \sqrt{\frac{4\pi n e^2h}{m\varkappa}},
\end{equation}
where $V_p$ is the velocity of gated plasmons and $\varkappa$ is the dielectric permittivity in the space between 2DES and the gate.

Initially observed in 2D systems of electrons on a liquid helium surface~\cite{Grimes1976} as well as in silicon inversion layers~\cite{Allen1977,Theis1977,Tsui1980}, 2D plasmons continue to be actively investigated in various 2D structures~\cite{Lusakowski2017,Kukushkin2003,Kukushkin2006collective,Muravev2011observation,Scalari2012,Dyer2013,Muravev2015,Muravev2015_rel,Grigelionis2015,Muravev2017,Muravev2020,Woessner2014,Iranzo2018,Bandurin2018,Bylinkin2019,Kumar2016}. It should also be mentioned that 2D plasmons, especially in structures with metal gates, have proven promising as detectors and emitters of radiation in the terahertz range~\cite{Dyakonov1993,Peralta2002,Aizin2006,Knap2009,Popov2011,Aizin2007,Dyer2012,Muravev2012,Sydoruk2015,Svintsov2018}.

Recently, a new type of plasmons, referred to as near-gate~\cite{Zabolotnykh2019} or proximity~\cite{Muravev2019} plasmons, was discovered in a system of an infinite 2DES with an ideal metal strip in its vicinity. Unlike the gated plasmons, the fundamental mode of the near-gate plasmon was found to have square-root dispersion, which is defined in the long-wavelength limit as:
\begin{equation}
\label{disp_fund}
	\omega_{ng}(q_y)=\sqrt{\frac{8\pi e^2 n h}{m\varkappa}\frac{|q_y|}{W}},
\end{equation}
where $\varkappa$ is the dielectric permittivity in the space between the gate and 2DES, $W$ is the width of the gate strip, and $q_y$ is the plasmon wave vector along the strip. Such a surprising spectrum results from substantial electrical currents flowing outside the gated area of 2DES despite the fact that the charge density of this mode is localized almost entirely under the gate. This plasmon mode has been observed experimentally in 2DESs based on GaAs/AlGaAs quantum wells with a strip-shaped gate~\cite{Muravev2019,Zarezin2020}. Near-gate plasmons have also been studied theoretically~\cite{Zabolotnykh2019_Disk} and experimentally~\cite{Muravev2019_Disk} in systems with a disk-shaped gate, showing good agreement between the theory and experiment. 

Given gated or ungated 2DES with imposed perpendicular magnetic field $\bm B$, the frequency of magnetoplasmons $\omega_{mp}(B)$ becomes:
\begin{equation}
\label{spectr_mp}
	\omega_{mp}(B)=\sqrt{\omega^2(B=0)+\omega_c^2} 
\end{equation} 
where $\omega(B=0)$ is the frequency of an ungated~(\ref{plasmon}), gated~(\ref{gated}), or near-gate~(\ref{disp_fund}) plasmon in the absence of magnetic field, and $\omega_c=|e|B/(mc)$ is the electron cyclotron frequency in 2DES. 

In this paper, we investigate near-gate plasmons under more realistic conditions. While in the work~\cite{Zabolotnykh2019}, the conductivity of the strip was assumed to be infinitely large (ideal metal), our goal here is to take into account the finiteness, as well as the frequency and magnetic field dependence of the strip conductivity. In fact, we consider the case of a partially bounded two-layer system, with the first layer --- an infinite 2DES, and the second layer --- an infinitely long strip of finite width.

Plasmons in double-layer systems with two infinite layers have been well-studied~\cite{Economou1969,Shevchenko1976, Lozovik1976,Eguiluz1975,Sarma1981,Vitlina1981,Sarma1982, Santoro1988,Flensberg1994,Sarma1998,Chaplik2015}. Such a system is known to support optical and acoustic plasmon modes. As for the charge carriers of the same sign, these two plasmon types correspond, respectively, to the in-phase and out-of-phase oscillations of the charges in the layers. It has also been established that the optical mode has square root dispersion $\omega\propto \sqrt{q}$ while the acoustic plasmon mode has linear dispersion $\omega\propto q$ and lower frequency.

In present work, we explore plasma excitations in a partially bounded two-layer system, looking for plasmons that are localized near and propagate along the strip. It should be noted that here we consider only the acoustic plasmon mode, corresponding to the near-gate plasmon in the limit of infinite strip conductivity. The study of the optical plasmon mode is beyond the scope of this paper. Nevertheless, in Sec.~\ref{Sec:Discussion}, we include some qualitative analysis addressing the subject.

Our investigation reveals that in the absence of the external magnetic field, making the strip conductivity finite leads to the softening of the square-root dispersion law of the fundamental near-gate plasmon mode (\ref{disp_fund}), which becomes linear in the long-wavelength limit. When the system is subject to a perpendicular magnetic field, the fundamental plasmon mode is preserved gapless, in contrast to the case of an ideal metal strip~\cite{Zabolotnykh2019}, where it has a frequency gap at zero wave vector~(\ref{spectr_mp}).

\section{Analytical approach and principal equations}\label{Sec:Approach}

\begin{figure}[!t]
			\includegraphics[width=7.0cm]{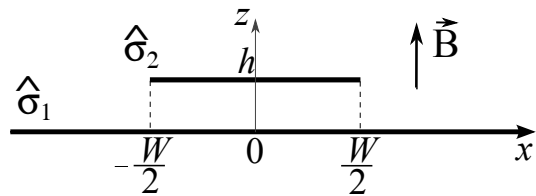}
			\caption{ \label{Fig:g} The two-layer system under consideration: the first layer is an infinite 2DES in $x-y$ plane; the second layer is a strip of width $W$, infinite in $y$-direction. Dynamical conductivity tensors of the first and second layers are $\hat{\sigma}_1$ and $\hat{\sigma}_2$, respectively.  
}
\end{figure}

In our analysis, we consider an infinite 2DES in $x-y$ plane at $z=0$ and a strip-shaped 2DES at $z=h$, infinite in $y$-direction. The strip has a finite width $W$ spanning the interval $[-W/2,W/2]$ in $x$-direction, as shown in Fig.~\ref{Fig:g}. The dielectric permittivity of the surrounding medium is $\varkappa$. The system is placed in a constant magnetic field $\bm B$, directed along the $z$-axis. The dynamical conductivity tensors of the infinite and strip-shaped 2DESs are given by $\sigma_{1,ij}(\omega)$ and $\sigma_{2,ij}(\omega)$, accordingly.

To determine the plasmon spectra, we follow the same approach used in~\cite{Zabolotnykh2019}. We look for the solutions in the form of waves, propagating along the strip, $\exp(iq_y y-i\omega t)$, and consider the spectra in the long-wavelength limit $|q_y|\ll k_F$, neglecting a spatial dispersion in conductivity tensors and the electromagnetic retardation effects.

Applying the Poisson equation together with the Ohm's law and continuity equation to the first layer, and then taking the Fourier transform, we obtain:
\begin{equation}
 \label{system_pot}
 	\begin{array}{lcr}
    \varphi_1(q_x)=\frac{2\pi}{\varkappa\sqrt{q_y^2+q_x^2}}\left(\rho_1(q_x)+\rho_2(q_x)e^{-h\sqrt{q_y^2+q_x^2}}\right),\\
   \varphi_2(q_x)=\frac{2\pi}{\varkappa\sqrt{q_y^2+q_x^2}}\left(\rho_1(q_x)e^{-h\sqrt{q_y^2+q_x^2}}+\rho_2(q_x)\right),\\
   i\omega\rho_1(q_x)=\sigma_{1,xx}(q_x^2+q_y^2)\varphi_1(q_x),
  \end{array}
\end{equation}
where $\varphi_1(x)$ and $\rho_1(x)$ are the plasmon potential and charge density in the first layer, $\varphi_2(x)$ and $\rho_2(x)$ are the potential and charge density in the strip, with $\rho_2(x)$ equal zero outside the strip, and argument $q_x$ denoting the respective Fourier transformation.

Eliminating $\varphi_1(q_x)$ and $\rho_1(q_x)$ in Eqs.~(\ref{system_pot}), and taking the inverse Fourier transform yields:
\begin{equation}
\label{exact_eq}
	\varphi_2(x)=\frac{1}{\varkappa}\int_{-\infty}^{+\infty}\frac{\varepsilon_{gated}(q,\omega)}{q\varepsilon_{2D}(q,\omega)}\rho_2(q_x)e^{iq_x x}dq_x,
\end{equation}
where  $q=\sqrt{q_x^2+q_y^2}$, and  
\begin{equation}
\label{diel_gated}
	\varepsilon_{gated}(q,\omega)=1-\frac{2\pi\sigma_{1,xx}q}{i\omega\varkappa}\left(1-e^{-2hq}\right)
\end{equation}
is the effective dielectric permittivity of the infinite gated 2DES. Here, in the special case of zero magnetic field and $qh\ll 1$, the condition $\varepsilon_{gated}(q,\omega)=0$ defines the spectrum of gated plasmons (\ref{gated}). In (\ref{exact_eq}), we also introduce the effective dynamical dielectric permittivity of infinite 2DES:
\begin{equation}
\label{diel_2D}
	\varepsilon_{2D}(q,\omega)=1-\frac{2\pi\sigma_{1,xx}q}{i\omega\varkappa}.
\end{equation}
We note that in the absence of magnetic field, the equation $\varepsilon_{2D}(q,\omega)=0$ defines an ordinary spectrum of 2D plasmons~(\ref{plasmon}).

To simplify the integral equation in (\ref{exact_eq}) and to find its analytical solution, we make two reasonable assumptions~\cite{Zabolotnykh2019}. First, we assume the separation distance $h$ between the 2DES layers to be small compared with the strip width $W$ and the characteristic length of the plasmon charge inhomogeneity $q^{-1}$, such that $qh\ll 1$ and $h/W\ll 1$. Then, in Eq.~(\ref{diel_gated}) we can make the following approximation:
\begin{equation}
\label{cond_h}
		\varepsilon_{gated}(q,\omega)\approx 1-\frac{4\pi\sigma_{xx}hq^2}{i\omega\varkappa}.
\end{equation}

The second assumption is that in the expression for the dynamical dielectric permittivity of 2DES~(\ref{diel_2D}), the second term dominates, i.e. $|2\pi \sigma_{1,xx}q_y/(i\omega\varkappa)| \gg 1$. Qualitatively, this means that for the given 2DES, the frequency of the plasmon under consideration, $\omega$, is small enough to become a major contributing factor in the dynamical dielectric permittivity, as well as, for instance, in the system response to an external alternating electric field. In the 'clean' limit of the Drude model for the conductivity tensor, when the electron relaxation time tends to infinity~(\ref{Drude}), this assumption can be formulated through the following inequalities:  
\begin{equation}
\label{cond_bulk}
	\omega^2\ll \omega_c^2+\omega_{p1}^2(q_y)\quad \text{at}\quad |\omega|>|\omega_c| 
\end{equation}
and
\begin{equation}
\label{cond_edge}
	\omega^2\gg \omega_c^2-\omega_{p1}^2(q_y)\quad \text{at} \quad|\omega|<|\omega_c|.
\end{equation}  
Here, $\omega_c=|e|B/(m_1c)$, $m_1$, and $\omega_{p1}$ designate, respectively, the electron cyclotron frequency, effective electron mass, and 2D plasmon frequency~(\ref{plasmon}) in the first layer. It should be noted that condition (\ref{cond_bulk}) implies the frequency of the given plasmon $\omega$ to be much lower than that of 2D magnetoplasmons~(\ref{spectr_mp}) excited in an infinite 2DES large distance away from the strip.

Based on the two assumptions above, we obtain the following approximation related to Eq.~(\ref{exact_eq}):
\begin{equation}
\label{fin_approx}
	\frac{\varepsilon_{gated}(q,\omega)}{q\varepsilon_{2D}(q,\omega)}\approx 2h-\frac{i\omega\varkappa}{2\pi\sigma_{1,xx}q^2}.
\end{equation}

Using this simplification, we rewrite the integral equation~(\ref{exact_eq}) as:
\begin{eqnarray}
\label{fin_int}
	\varphi_2(x)=\frac{4\pi h}{\varkappa}\rho_2(x)\nonumber\qquad\qquad&\\ -\frac{i\omega}{2\sigma_{1,xx}|q_y|}\int^{W/2}_{-W/2} e^{- |q_y| | x- x'|} \rho_2( x')d x'.&
\end{eqnarray}

To bring this equation into a more convenient form, we can express $\varphi_2(x)$ in terms of $\rho_2(x)$. To do so, we determine the relationship between $\varphi_2(x)$ and $\rho_2(x)$ from the material equations for the strip. Thus, applying the Ohm's law and the continuity equation, we obtain:
\begin{equation}
\label{c2}
	\left(\partial_x^2-q_y^2\right)\varphi_2(x)= \frac{-i\omega}{\sigma_{2,xx}}\rho_2(x).
\end{equation}

Next, we impose standard boundary conditions for the potential $\varphi_2(x)$ --- vanishing normal component of the current density $\bm j_2(x)$ at the edges of the strip:
\begin{equation}
\label{zero_cur}
	j_{2,x}(\pm W/2)=- \left(\sigma_{2,xx}\partial_x+\sigma_{2,xy}iq_y\right)\varphi_2(x)|_{x=\pm W/2}=0.
\end{equation}

Given Eq.~(\ref{c2}) and the boundary conditions (\ref{zero_cur}), $\varphi_2(x)$ can be expressed in terms of $\rho_2(x)$ using Green's function $G(x,x')$ defined as:
\begin{equation}
	(\partial_x^2-q_y^2)G(x,x')=\delta(x-x'),
\end{equation}
where $G(x,x')$ should satisfy the boundary conditions in (\ref{zero_cur}).

We find $G(x,x')$ to be of the following form (see also~\cite{Cataudella1987,Rudin1997}):
\begin{eqnarray}
	G(x,x')=-\frac{b\exp(|q_y|(x+x'))+b^{-1}\exp(-|q_y|(x+x'))}{4|q_y|\sinh(|q_y|W)} \nonumber &\\
	+\frac{2\cosh(|q_y|W-|q_y||x-x'|)}{4|q_y|\sinh(|q_y|W)},\,\qquad&
\end{eqnarray}
where parameter $b$ is defined as:
\begin{equation}
	\label{b}
	b=\frac{\sigma_{2,xx}-i\sigma_{2,xy}sgn(q_y)}{\sigma_{2,xx}+ i\sigma_{2,xy}sgn(q_y)}.
\end{equation}
Now we can rewrite the first term in (\ref{fin_int}) as:
\begin{equation}
	\label{pot2}
	\varphi_2(x)=\frac{-i\omega}{\sigma_{2,xx}}\int^{W/2}_{-W/2}G(x,x') \rho_2(x')dx',
\end{equation}
where $-W/2\le x \le W/2$.

Finally, substitution of (\ref{pot2}) into (\ref{fin_int}) yields a single integral equation for the charge density in the strip $\rho_2(x)$:
\begin{eqnarray}
\label{int_fin}
		\frac{4\pi h}{\varkappa} \rho_2(x)= \frac{i\omega}{2\sigma_{1,xx}|q_y|}\int^{W/2}_{-W/2} e^{- |q_y| | x- x'|} \rho_2( x')d x'\nonumber &\\ +\frac{-i\omega}{\sigma_{2,xx}}\int^{W/2}_{-W/2}G(x,x')\rho_2(x')dx'.\qquad&
\end{eqnarray}
Since this integral equation has exponential kernels, it can be reduced to a differential equation with some boundary conditions~\cite{Polyanin2008}. Hence, we can reduce Eq.~(\ref{int_fin}) to the differential equation valid for $-W/2<x<W/2$:
\begin{equation}
\label{scr}
	\left(\partial_x^2-q_y^2+\frac{i\omega\varkappa}{4\pi h}\left(\frac{1}{\sigma_{1,xx}}+\frac{1}{\sigma_{2,xx}}\right)\right)\rho_2(x)=0
\end{equation}
with the boundary conditions at the strip edges defined as:
\begin{eqnarray}
\label{bc}
\left(\partial_x+\frac{|q_y|C_{\pm}}{D}\right)\rho_2(x) |_{x=\pm W/2} \mp  
	\frac{|q_y| F} {D} \rho_2\left(\mp W/2\right)=0,\nonumber\\
	\text{where} \quad C_{\pm}=\pm(1-e^{-2|q_y|W})\sigma_{1,xx}^{-2}\qquad\qquad\qquad\quad\\ 
	+(\pm 2-b+b^{-1})\sigma_{1,xx}^{-1}\sigma_{2,xx}^{-1}+(b^{-1}-b)\sigma_{2,xx}^{-2},\quad\nonumber\\
	D=\left(\sigma_{1,xx}^{-1}+\sigma_{2,xx}^{-1}(1+b)\right)\left(\sigma_{1,xx}^{-1}+\sigma_{2,xx}^{-1}(1+b^{-1})\right)\nonumber\\
	-\sigma_{1,xx}^{-2}e^{-2|q_y|W},\qquad\nonumber\\
	F=2e^{-|q_y|W}\sigma_{1,xx}^{-1}\sigma_{2,xx}^{-1},\qquad\qquad\qquad\qquad\nonumber
\end{eqnarray}
and the upper and lower signs correspond to the first and second condition.

At this point, let us briefly analyze the asymptotics of the resultant equation and boundary conditions. Formal substitution of the partial derivative $\partial_x$ in Eq.~(\ref{scr}) by $i q_x$ yields the dispersion equation for acoustic plasmons in an infinite two-layer system \cite{Economou1969,Sarma1981,Sarma1982}. In the limit of infinite conductivity $\sigma_{2,xx}$ or $\sigma_{1,xx}$, we obtain the expression describing gated plasmons~\cite{Chaplik1972} in the first or second infinite layer, accordingly. As for the boundary conditions (\ref{bc}), in the limit of infinite strip conductivity, $\hat{\sigma}_2\to \infty$, we have $C_{\pm}/D\to \pm 1$ and $F/D\to 0$, which leads to the previously derived boundary conditions for the case of the ideal metal strip~\cite{Zabolotnykh2019}. In the limit of $\hat{\sigma}_1\to \infty$, we arrive at the case of the gated strip, where $C_{\pm}/D\to sgn(q_y)\sigma_{2,xy}/\sigma_{2,xx}$, $F/D\to 0$, and the boundary conditions at $x=\pm W/2$ become $(\sigma_{2,xx}\partial_x+i q_y \sigma_{2,xy})\rho_2(x)=0$, which corresponds to the normal component of the current density vanishing at the boundary, $j_{2,x}(\pm W/2)=0$, since in the gated system $\rho_2(x)\propto \varphi_2(x)$.

It should be mentioned that Eqs.~(\ref{scr}) and (\ref{bc}) are derived for an arbitrary 2D conductivity tensor. Further on in this paper, we consider only the Drude model for the conductivity tensor, assuming the electron relaxation time in 2DESs to be infinite~(\ref{Drude}).

In the following Sections~\ref{spectrum_wo} and \ref{spectrum_in}, we analyze given plasmon spectra obtained in the absence and in the presence of the external magnetic field.


\section{Plasmon spectra with no external magnetic field}\label{spectrum_wo}
First, let us consider plasmons at zero magentic field. In this case, $\sigma_{xx}=\sigma$, $\sigma_{xy}=0$, and $b=1$, see Eq. (\ref{b}). Now, the solutions to Eqs.~(\ref{scr}) and (\ref{bc}) have certain parity across the strip, thus, requiring only one boundary condition:
\begin{equation}
\label{bc_zmf}
	\left(\partial_x+|q_y|\left(1+\frac{2\sigma_1}{\sigma_{2}(1\mp e^{-|q_y|W})}\right)^{-1}\right)\rho_2(x)|_{x=W/2}=0,
\end{equation}
where the $''-''$ and $''+''$ signs refer to the even and odd modes, accordingly.

In the following discussion, we use the Drude model~(\ref{Drude}) at zero magnetic field to analyze conductivities of the first and second layers, $\sigma_{1}$ and $\sigma_{2}$, denoting the respective electron concentrations by $n_1$ and $n_2$, and effective masses by $m_1$ and $m_2$. In the given case, the even and odd solutions of Eq.~(\ref{scr}) are, correspondingly, $\cos kx$ and $\sin kx$, with the effective transverse wave vector $k$ defined as:
\begin{equation}
		k^2=\frac{\omega^2}{V_a^2}-q_y^2,
\end{equation}
where $V_a^{-2}=V_{p1}^{-2}+V_{p2}^{-2}$ is the velocity of acoustic plasmons in a two-layer system, and $V_{p1,2}^2=4\pi e^2n_{1,2}h/(\varkappa m_{1,2})$.

Applying the boundary conditions in (\ref{bc_zmf}), we arrive at the dispersion relation below:
\begin{equation}
\label{dr_zmf}
	k\left(\tan \frac{kW}{2}\right)^{\pm 1}=\pm |q_y|\left(1+\frac{n_1 m_2}{n_2 m_1}\frac{2}{1\mp e^{-|q_y|W}}\right)^{-1},
\end{equation}
where the upper and lower signs indicate the even and odd modes, respectively.
\begin{figure}[!t]
			\includegraphics[width=8.0cm]{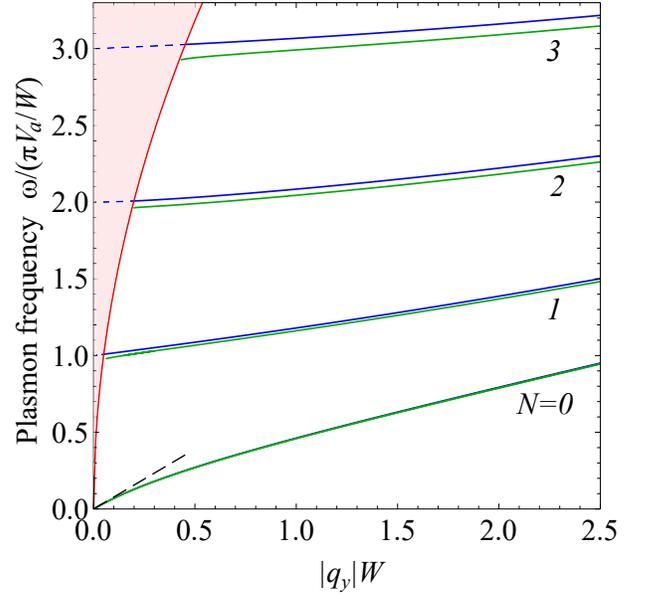}
			\caption{ \label{Fig:Spectrum} Spectra of plasmons in a partially bounded two-layer system (see Fig.~\ref{Fig:g}), in the absence of the external magnetic field.  Blue and green lines denote data for $N=0,1,2,3$ modes obtained by analytical and numerical methods, respectively. Red-shaded area designates the region $\omega>\omega_{p1}(q_y)$~(\ref{plasmon}) corresponding to the continuum of 2D plasmons in the first layer. Within and in the vicinity of this region, our analytical results become less accurate --- see the discussion preceding Eqs.~(\ref{cond_bulk}) and (\ref{cond_edge}). The dashed line represents the long-wavelength asymptote~(\ref{fund_wo}). For the computed data, $h/W=0.003$ and $n_1m_2/(n_2 m_1)=1/5$.  
}
\end{figure}
From Eq.~(\ref{dr_zmf}) we obtain a discrete series of plasmon modes with frequencies $\omega_N(q_y)$, where $N=0,1,2,...$ designates the number of nodes in charge density across the strip. The spectra for the first four modes of $\omega_N(q_y)$ are plotted in Fig.~\ref{Fig:Spectrum}. In the long-wavelength limit, when $|q_y|W\ll 2n_1 m_2/(n_2 m_1)$ and $|q_y|W\ll 1$, the fundamental mode $N=0$ has linear dispersion as follows:
\begin{equation} 
\label{fund_wo}
	\omega_{N=0}(q_y)=V_{p2}|q_y|. 
\end{equation}
As the conductivity of the strip tends to infinity, i.e. $n_2 \to \infty$ (ideal metal strip), the $q_y$ interval of linear dispersion vanishes, and we arrive to the square-root plasmon spectrum of the fundamental mode for $|q_y|W\ll 1$ derived in the previous paper~\cite{Zabolotnykh2019}. Here, we note the fundamental mode frequency to be lower than that of 2D plasmons in the first layer. Therefore, the condition in~(\ref{cond_bulk}) is satisfied in the absence of magnetic field. 

In addition, we determine the spectrum $\omega_N(q_y)$ numerically for the wave vectors and frequencies outside the continuum of 2D plasmons in the first layer, i.e. for $\omega_N(q_y)<\omega_{p1}(q_y)$. In the exact Eqs. (\ref{exact_eq}) and (\ref{pot2}), we expand $\rho_g(x)$ into the series of $\sin(\pi P x/W)$, for $P=1,3,5,..$, and $\cos(\pi P x/W)$, for $P=0,2,4,..$, to find the odd and even modes, respectively. Then, following a standard computational procedure, we arrive at the spectra plotted in green in Fig.~\ref{Fig:Spectrum}. For the fundamental mode $N=0$, the numerical and analytical solutions match perfectly. For the higher excited modes, $N=1,2,3,..$, the results are in good agreement overall, although numerical solution yields slightly lower frequencies.

For small values of the wave vector $q_y$, higher modes, with $N\ge 1$, fall inside the bulk continuum of plasmons in infinite 2DES. In this region, the analytically obtained spectra become less accurate, as the condition~(\ref{cond_bulk}) is not satisfied. Nevertheless, for a qualitative insight into plasmon spectra, we include the asymptotics for small $q_y$ as follows. For even modes $N=2,4,..$, we find $\omega_N/V_a=\pi N/W+q_y^2 W(n_2m_1/(n_1m_2)+1/2)/(\pi N)$ for $|q_y| W\ll 2n_1m_2/(n_2m_1)$ and $|q_y| W\ll 1$. For odd modes $N=1,3,..$, we find $\omega_N/V_a=\pi N/W+2|q_y|/(\pi N(1+n_1m_2/(n_2m_1)))$ for $|q_y| W\ll 1$.

As $|q_y|\to \infty$, all the modes exhibit asymptotic behavior described by $\omega_N^2/V_a^2=\pi^2(N+1)^2/W^2+q_y^2$, i.e. $\omega_N$ approaches the frequency of the acoustic plasmon mode in a two-layer system, with $q_x \rightarrow \pi(N+1)/W$.


\section{Plasmon spectra in the presence of external magnetic field}\label{spectrum_in}

Next, we consider the system placed in a perpendicular constant magnetic field $\bm B$ (Fig.~\ref{Fig:g}), with conductivity tensors of the given layers described by the Drude model~(\ref{Drude}). Here, for simplicity, we assume the effective mass in both 2DESs to be the same, i.e. $m_1=m_2=m$. Consequently, the difference between $\hat{\sigma}_1$ and $\hat{\sigma}_2$ can be due only to dissimilar electron concentrations in the layers, $n_1$ and $n_2$.   
Note also that below we assume $\omega\neq \omega_c$ since we are considering collisionless limit of the Drude model~(\ref{Drude}).

\begin{figure}[!t]
	\includegraphics[width=8.0cm]{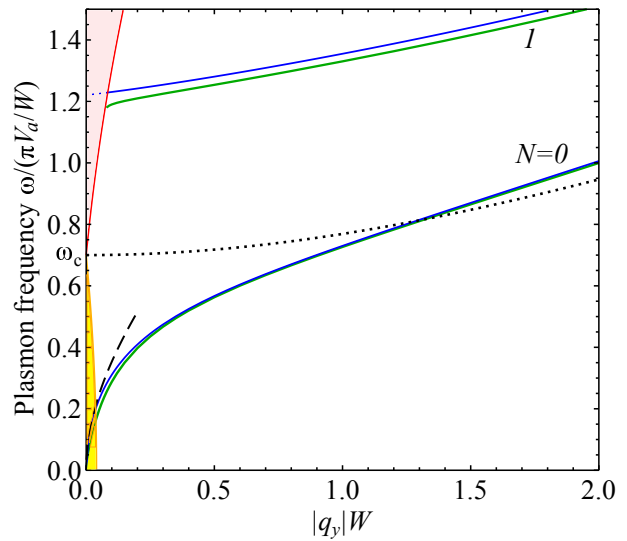}
	\caption{ \label{Fig:Spectrum_MF} Spectra of the plasmons in a partially bounded two-layer system subject to an external magnetic field (Fig.~\ref{Fig:g}). Analytically and numerically obtained data are plotted in the blue and green lines, correspondingly. The dashed line denotes the long-wavelength asymptote~(\ref{mf_as}). The red- and yellow-shaded areas, designate the regions defined by inequalities $\omega^2>\omega_c^2+\omega_{p1}(q_y)$ and $\omega^2<\omega_c^2-\omega_{p1}(q_y)$, respectively. Within and in the vicinity of the two regions, our approach results in appreciable discrepancy --- see the discussion preceding Eqs.~(\ref{cond_bulk}) and (\ref{cond_edge}). For comparison, the dotted line represents the dispersion law of acoustic plasmons in an infinite two-layer system, $\omega^2=\omega_c^2+V_a^2q_y^2$. For the computed data, $h/W=0.005$, $n_1/n_2=1/5$, $\omega_c/(\pi V_a/W)=0.7$.
	}
\end{figure}

Unlike the previous case, in the presence of the magnetic field, solutions for the charge density $\rho_2(x)$ no longer have parity across the strip, as boundary conditions lead to the intermixture of even and odd solutions. Therefore, we look for $\rho_2(x)$ in the form of a linear combination of $\sin kx$ and $\cos kx$, where $k=\sqrt{(\omega^2-\omega_c^2)/V_a^2-q_y^2}$ corresponds to the effective wave vector across the strip. To find the plasmon spectra, we substitute $\rho_2(x)$ into the boundary conditions in (\ref{bc}) and then derive the analytical dispersion equation. As this leads to a fairly cumbersome expression, we do not include it here. The resultant characteristic plasmon spectrum is shown in Fig.~\ref{Fig:Spectrum_MF}, where the blue and green curves refer to the analytically and numerically obtained solutions, accordingly. Clearly, the outcomes of both methods indicate close agreement.

In the figure, the red- and yellow-shaded regions are defined by respective inequalities: $\omega^2>\omega_c^2+\omega_{p1}(q_y)$ and $\omega^2<\omega_c^2-\omega_{p1}(q_y)$. Here, our analytical solution formally becomes inappropriate, according to Eqs.~(\ref{cond_bulk}) and (\ref{cond_edge}). Nonetheless, it is important to emphasize the difference between these two areas. In the red-shaded zone, $\varepsilon_{2D}(q,\omega)$~(\ref{diel_2D}) becomes zero for a certain $q_x$,  corresponding to the excitation of 2D magnetoplasmons in the first layer. Therefore, even if plasmon modes localized near the strip exist, they strongly fade due to their interaction with 2D magnetoplasmons in this layer. On the other hand, in the yellow-shaded region, $\varepsilon_{2D}(q,\omega)$ does not go to zero. Here, our method is inaccurate to some degree, as the unity in Eq.~(\ref{diel_2D}) can no longer be neglected when carrying out the integration in Eqs.~(\ref{exact_eq}) for small values of $q_x$. However, the localized plasmon mode still exists in this zone, while our analytical solution within and in the vicinity of this area merely shows a slight deviation from the actual dispersion curve obtained by a numerical method, as can be seen in Fig.~\ref{Fig:Spectrum_MF}.  
Although for small $q_y$ the fundamental mode $N=0$ lies inside the yellow-shaded area, for its qualitative description, we find the asymptotic behavior at $|q_y|W\ll 1$, $|\omega/\omega_c| \ll 1$ and $V_a|q_y| \ll |\omega_c|$ to be as follows:
\begin{eqnarray}
\label{mf_as}
	\omega^2_{N=0}=\frac{\omega_c^2|q_y|Wn_2}{2(1+n_1/n_2)n_1} \nonumber
	+ \frac{|\omega_c q_y|V_a}{1+n_1/n_2}\tanh\frac{|\omega_c|W}{2V_a}+ &\\
	+\frac{q_y^2WV_a n_2}{2n_1}|\omega_c|\coth\frac{|\omega_c|W}{2V_a}+q_y^2V_a^2. \qquad
\end{eqnarray}
In the yellow-shaded region, the numerically obtained solution for the fundamental mode seems to indicate a linear trend, rather than analytically derived square-root dependency at $q_y \to 0$ (\ref{mf_as}). Importantly, both solutions show gapless behavior of the fundamental mode, in contrast to the case of near-gate plasmons~\cite{Zabolotnykh2019}, which at zero wave vector exhibit a frequency gap equal to the cyclotron frequency~(\ref{spectr_mp}). We also note that in part, the fundamental mode lies below the dispersion of plasmons in an infinite two-layer system (the dotted line in Fig.~\ref{Fig:Spectrum_MF}), meaning that in this spectral area, $k$ becomes purely imaginary and the plasmon charge density $\rho_2(x)$ tends to be localized near the strip edges.

As for the higher modes $N\ge 1$, the analytically obtained spectra become inappropriate at small values of  $q_y$ because of the violated condition in~(\ref{cond_bulk}). However, we still include the asymptotics that are valid for the moderate values of $q_y$, when conditions $|q_y|W\ll 1$, $|q_y|W\ll 2n_1/n_2$, $|q_y|V_a\ll \omega_c$ and~(\ref{cond_bulk}) are satisfied. Thus, for even modes $N=2,4,..$, we find: 
\begin{equation}
\label{as_mf_ev}
	\omega_N(q_y)=\omega_N(0)+\frac{q_y^2V_a^2}{\omega_N(0)}\left(\frac{1}{2}+ \frac{n_2\pi^2V_a^2N^2}{n_1\omega_N^2(0)W^2}+\frac{2\omega_c^2}{\pi N \omega_N^2(0)}\right);\nonumber
\end{equation}
while for odd modes $N=1,3,..$, we find: 
\begin{equation}
\label{as_mf_odd}
	\omega_N(q_y)=\omega_N(0)+\frac{2|q_y|V_a^2}{W(1+n_1/n_2)}\frac{(\pi V_a N/W)^2}{\omega_N^3(0)};
\end{equation}
where $\omega_N(0)=\sqrt{(\pi V_a N/W)^2+\omega_c^2}$ is the frequency at $q_y=0$. As $|q_y|\to \infty$, all the modes follow standard asymptotic behavior described by: $\omega_N^2(q_y)=\omega_c^2+V_a^2\left[\pi^2(N+1)^2/W^2+q_y^2\right]$.

\begin{figure}
			\includegraphics[width=8.0cm]{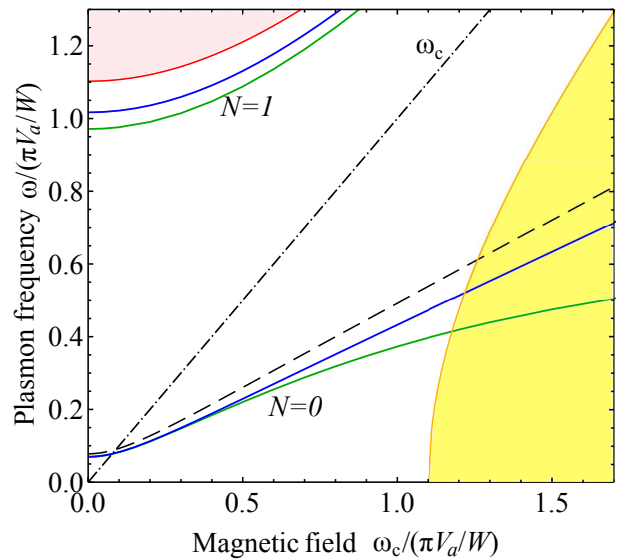}
			\caption{ \label{Fig:MD} Magnetodispersion of plasmons under study. Analytical and numerical results are plotted in the blue and green lines, correspondingly. The dashed line marks the long-wavelength asymptote~(\ref{mf_as}). The red- and yellow-shaded regions are analogous to those in Fig.~\ref{Fig:Spectrum_MF}. The dash-dotted line is the 'cyclotron frequency' $\omega=\omega_c$. For the computed data, $h/W=0.005$, $n_1/n_2=1/5$, and $q_y W=0.1$.
}
\end{figure}

Last but not least, we explore the magnetodispersion of the plasmons under study, i.e. the dependence of plasmon frequency on the magnitude of the magnetic field, which is often measured experimentally. In Fig.~\ref{Fig:MD}, we include the characteristic magnetodispersion computed for $|q_y|W=0.1\ll 1$. Here, the higher modes $N=1,2,...$ exhibit ordinary magnetodispersion described by Eqs.~(\ref{as_mf_ev}) and (\ref{as_mf_odd}), when the plasmon frequency tends to the electron cyclotron frequency with increasing magnetic field. In contrast, the fundamental mode $N=0$ indicates quite a non-trivial dependency. For its qualitative interpretation, we consider two special limiting cases --- plasmons in the gated strip with $n_1/n_2\to \infty$, and near-gate plasmons with $n_1/n_2\to 0$. In the first instance, plasma excitations show no magnetodispersion at $qh \ll 1$ (similar to the case of gated edge magnetoplasmons~\cite{Volkov1988,Nazin1987}), i.e. their frequency is independent of the magnetic field. At the opposite extreme, however, we have typical magnetodispersion for near-gate plasmons defined in~(\ref{spectr_mp}). Thus, at finite values of $n_1/n_2$, we expect to see some degree of magnetodispersion, ranging in between the given limiting cases. Indeed, this qualitative consideration has been confirmed by numerical and analytical calculations, as shown in Fig.~\ref{Fig:MD}. In addition, from~(\ref{mf_as}) it follows that with an increase in magnetic field (but outside the yellow-shaded region and if $n_1/n_2$ is of the order of unity), the plasmon frequency approaches its asymptotic value:
\begin{equation}
 \omega\approx \omega_c\sqrt{\frac{|q_y|Wn_2}{2n_1(1+n_1/n_2)}}.
\end{equation}

\section{Discussion and conclusion}\label{Sec:Discussion}
In this paper, we focus on the acoustic plasmon mode in a two-layer electron system, where the charges in the infinite 2DES and the strip oscillate out-of-phase, partially 'screening' each other. However, as we have already mentioned, along with this type of plasmon, there should exist the optical plasmon mode, with charges in the two layers oscillating in phase. Although the analysis of this mode is beyond the scope of present work, here we include a related qualitative description. Considering the optical plasmon mode at small values of separation distance $h$ between the infinite 2DES and the strip, i.e. under the conditions of $|q_y|h\ll 1$ and $h/W \ll 1$, the given two-layer system can be treated as a single 2DES with inhomogeneous conductivity --- $\sigma_1$ for $|x|>W/2$ and $\sigma_1+\sigma_2$ for $|x|<W/2$. It has been established that in such a system, near $x=\pm W/2$, there exist so-called inter-edge magnetoplasmons~\cite{Mikhailov1992,Mikhailov1995}. Therefore, it is likely that optical plasmon modes correspond to these inter-edge magnetoplasmons excited at $x=\pm W/2$, provided the strip width $W$ is sufficiently large to prevent their interaction. Otherwise, optical plasmon modes can be regarded as the result of the hybridization of these excitations.

In our analysis we neglected the electromagnetic retardation effects, therefore the obtained results are applicable only for $\omega \ll cq/\sqrt{\varkappa}$, where $c$ is the speed of light in vacuum. 

Thus far, we consider plasmons using collisionless Drude model (\ref{Drude}). Given finite relaxation time $\tau$ (for simplicity the same in both layers), plasmons under study exist if $\omega\tau\gg 1$ and strongly fade if $\omega\tau\ll 1$. If we extract $\tau$ from typical electron mobility for GaAs/AlGaAs quantum wells $\mu=5 \cdot 10^6$ cm$^2/$(V\,s) at 1.5 K \cite{Muravev2019}, then we find that the plasmons are well-defined at frequencies $\omega/(2\pi)$ larger than $1/(2\pi\tau)\approx 0.9$ GHz; experiments under consideration \cite{Muravev2019, Zarezin2020} are usually conducted at higher frequencies, which are of the order of $10$--$100$ GHz.

Let us also discuss qualitatively the consequences of taking into account the spatial dispersion in conductivity, which we neglected by using Drude model. Firstly, we disregard the ''electronic pressure'' contribution, which is characterized by the effective velocity $s$, where $s^2=v_F^2/2$ and $v_F$ is the Fermi velocity, see, for example, Eqs.~(53)--(57b) from Ref.~\cite{Fetter1973}. This contribution results in the additional term $s^2q^2$ in the squared plasmon frequency, thus it can be neglected at $\omega^2\gg s^2q^2$. Consider this condition in the case of the fundamental mode $N=0$. In the absence of magnetic field and at $|q_y| W\ll 1$ we have 	$\omega_{N=0}^2(q_y)=V_{p2}^2q_y^2$ (\ref{fund_wo}), so we arrive to the condition $V_{p2}^2\gg s^2$. If we consider for qualitative estimation two identical layers based on GaAs/AlGaAs quantum wells, with electron concentration $n=3 \cdot 10^{11}$ 1/cm$^2$ and the distance between the layers $h=440$ nm, which correspond to the experimental set up in the case of previously studied near-gate plasmons~\cite{Muravev2019,Zarezin2020}, then we obtain $V_{p2}^2/s^2\approx 170$, so electronic pressure is negligible. However, it should be taken into account for smaller separation distances, when $h$ is of the order of 20 nm or less. In the presence of perpendicular magnetic field another contribution to the spatial dispersion of conductivity arises. This contribution is due to the existence of so-called Bernstein magnetoplasma modes, see Refs.~\cite{Bernstein1958, Chiu1974}. Qualitatively, this contribution can be neglected if the plasmon frequency does not fall into frequency gaps, which are situated near harmonics of the cyclotron frequency $2\omega_c$, $3\omega_c$, ..., see Fig.~2 from Ref.~\cite{Chiu1974}, and when the long wavelength limit $qR_c\ll 1$ takes place, where $R_c$ is the electron cyclotron radius.

We reiterate that Eqs.~(\ref{scr}) and (\ref{bc}) are derived for an arbitrary conductivity model based on the approximation in~(\ref{fin_approx}), as was mentioned in the discussion preceding and following Eq.~(\ref{cond_h}). Besides the Drude model used in this paper, our method can be applied to describe plasmons in systems with other conductivity models, such as 2D conductivity in a strong magnetic field and quantum Hall regime, graphene conductivity, etc.

In summary, we have studied analytically as well as numerically acoustic plasmon modes in a partially bounded two-layer system comprised of an infinite 2DES and an infinite strip. The obtained plasmon spectra are characterized by the mode number $N$ and the wave vector $q_y$ directed along the strip. The fundamental mode $N=0$ is found to be gapless, whereas higher modes $N=1,2,...$ exhibit gapped dispersion law. Without the external magnetic field, the fundamental mode has linear dispersion in the long-wavelength limit, in contrast to the square-root dispersion of this mode established in the previously studied case of infinite strip conductivity (ideal metal strip)~\cite{Zabolotnykh2019}. In the presence of a perpendicular magnetic field, the fundamental mode retains gapless dispersion and shows non-trivial magnetodispersion. The magnetodispersion is found to be strongly affected by the relation between electron concentrations in the first and second layers.

\begin{acknowledgments}
	We would like to thank I. V. Kukushkin and V. M. Muravev for numerous stimulating discussions. The work was supported by the Russian Science Foundation (project no. 16-12-10411).
\end{acknowledgments}

\appendix*
\section{Drude model}
Consider a 'clean' 2DES, with infinite electron relaxation time, exposed to the constant magnetic field $\bm{B}$ applied perpendicular to the 2DES plane. Then, in the framework of Drude model, the dynamical longitudinal and transverse 2D conductivities, $\sigma_{xx}$ and $\sigma_{xy}$, can be expressed as:
\begin{equation}
\label{Drude}	
	\sigma_{xx}	=\frac{e^2n}{m}\,\frac{-i\omega}{-\omega^2+\omega_c^2},\quad \sigma_{xy}=\frac{e^2n}{m}\,\frac{-\omega_c}{-\omega^2+\omega_c^2},
\end{equation}
where $n$ is the electron concentration in 2DES, $-e$ and $m$ are the electron charge and effective mass, and $\omega_c=|e|B/(mc)$ is the electron cyclotron frequency in 2DES. Note, that $\omega \neq \omega_c$ as we consider 'clean' limit. Mention also that when deriving expressions~(\ref{Drude}), we neglect the spatial dispersion of conductivity.

\end{document}